\documentclass[review]{elsarticle}
\usepackage{lineno,hyperref}
\biboptions{authoryear}

\usepackage{placeins}
\usepackage{graphics}
\usepackage{wrapfig}

\usepackage{multicol}
\usepackage{amsmath,amsfonts,amssymb}
\usepackage{graphicx}
\usepackage{setspace}
\usepackage{mathtools}
\usepackage{scrextend}
\usepackage{hhline}
\usepackage{tabularx}
\usepackage[utf8]{inputenc}
\usepackage{multirow}
\usepackage{mathrsfs}
\usepackage{adjustbox}
\usepackage{bbm}
\usepackage{tablefootnote}
\usepackage{algorithm,algorithmic}
\usepackage{subfigure}

\usepackage{times}
\usepackage{mathptmx}
\usepackage[T1]{fontenc}
\usepackage[binary-units=true]{siunitx}
\usepackage{url}
\usepackage{xspace}
\usepackage{pgfplots}
\usepgfplotslibrary{groupplots}
\usepackage{csquotes}
\usepackage{tabto}
\usepackage[absolute,overlay]{textpos}
\usepackage{environ}
\makeatletter
\usepackage{mathptmx}

\usepackage{enumitem}
\usepackage{amsthm}

\usepackage{xcolor}

\usepackage{booktabs}
\usepackage{float}
\def\arrvline{\hfil\kern\arraycolsep\vline\kern-\arraycolsep\hfilneg}
\usepackage{scalerel,stackengine}

\usepackage{nomencl}
\usepackage{etoolbox}
\renewcommand\nomgroup[1]{%
  \item[\bfseries
  \ifstrequal{#1}{A}{Sets and indices}{%
  \ifstrequal{#1}{B}{Parameters}{%
  \ifstrequal{#1}{C}{Variables}{}}}%
]}

\usepackage{comment}

\usepackage{lipsum}
\makeatletter
\def\ps@pprintTitle{%
	\let\@oddhead\@empty
	\let\@evenhead\@empty
	\def\@oddfoot{}%
	\let\@evenfoot\@oddfoot}
\makeatother

\begin{document}
\begin{frontmatter}

\title{Enhancing Energy System Models Using Better Load Forecasts}
%\tnotetext[mytitlenote]{Fully documented templates are available in the elsarticle package on %\href{http://www.ctan.org/tex-archive/macros/latex/contrib/elsarticle}{CTAN}.}

%\author{Thomas Möbius\fnref{mail1}, Mira Watermeyer\fnref{mail2}, Oliver Grothe, Felix Müsgens}
%\address{BTU Cottbus-Senftenberg, Siemens-Halske-Ring 13, 03044 Cottbus}
%\fntext[mail1]{thomas.moebius@b-tu.de}
%\fntext[mail2]{mira.watermeyer@kit.edu}

\author[mainaddress1]{Thomas Möbius\corref{mycorrespondingauthor}}
\ead{thomas.moebius@b-tu.de}

\author[mainaddress2]{Mira Watermeyer\corref{mycorrespondingauthor}}
\cortext[mycorrespondingauthor]{Corresponding authors}
\ead{mira.watermeyer@kit.edu}

\author[mainaddress2]{Oliver Grothe}
\author[mainaddress1]{Felix Müsgens}

\address[mainaddress1]{BTU Cottbus-Senftenberg, Siemens-Halske-Ring 13, Cottbus}
\address[mainaddress2]{Karlsruhe Institute of Technology, Kaiserstraße 12, Karlsruhe}
%\address[mysecondaryaddress1]{}

\begin{abstract}

Energy system models require a large amount of technical and economic data, the quality of which significantly influences the reliability of the results. Some of the variables on the important data source ENTSO-E transparency platform, such as transmission system operators' day-ahead load forecasts, are known to be biased. These biases and high errors affect the quality of energy system models. 
We propose a simple time series model that does not require any input variables other than the load forecast history to significantly improve the transmission system operators' load forecast data on the ENTSO-E transparency platform in real-time, i.e., we successively improve each incoming data point. 
We further present an energy system model developed specifically for the short-term day-ahead market. We show that the improved load data as inputs reduce pricing errors of the model, with strong reductions particularly in times when prices are high and the market is tight. 

\end{abstract} 

\begin{keyword}
\texttt{Data preprocessing\sep Day-ahead electricity prices\sep Energy system modelling}
%\MSC[2010] 00-01\sep  99-00
\end{keyword}

%\date{\today}
%\pubMonth{May}
%\pubYear{2022}
%\pubVolume{Vol}
%\pubIssue{Issue}
%\JEL{}
%\Keywords{}

\end{frontmatter}

%\linenumbers

%\section{mögliche Titel}
%The Value of better Load Forecasts for Energy System Models

%Enhancing the performance of Energy System Models using better Load Forecasts

\section{Introduction} 
\label{intro}
Energy markets are complex and exhibit non-trivial interdependencies, so decisions from policy and industry stakeholders rely on theoretical models and other methodological support. Techno-economic energy system models are an essential tool, as they explain actual developments and provide valuable insights for future developments based on market fundamentals on the supply and demand side. They are capable of reflecting structural breaks better than most other model types. However, they rely on the quality of input data to provide accurate results. Literature has shown that important input data sets for energy systems models, in particular load data and wind or solar forecasts from official sources, often have significant systematic errors (\cite{Maciejowska2021, HIRTH2018}). These errors can be reduced in a preprocessing step, where they are considered as an econometric time series. This preprocessing step can exploit serial structures and predict future errors, which significantly reduces the errors. We analyse whether using these improved input data in an energy systems model will improve results. 

The contribution of this paper is threefold. First, we develop and provide a simple time-series model reducing forecast errors of hourly day-ahead load predictions of transmission system operators (TSOs) in real-time. We focus on load forecasts because they are the most correlated with the prices of the day-ahead electricity market and have the most potential for improvement compared with wind and PV forecasts (see, e.g., \cite{Maciejowska2021}). 
One advantage of our approach is that we take publicly available TSO-based load forecasts as given and thus, in modelling directly their prediction error as a predictable subject, do not need to develop a complex load forecast model. On country level, load forecasts are often used to represent the demand on the day-ahead market clearing.\footnote{In all main European markets, wholesale electricity prices are determined in the day-ahead market clearing one day before the actual delivery.} Thus, load forecasts are central variables for determining equilibria of demand and supply in energy system models. 

Second, we present a fundamental energy system dispatch model called the \emph{em.power dispatch} model, developed and calibrated precisely for short-term use in the day-ahead market. A primary objective of this model is to predict wholesale electricity prices. 
Using a rolling window, it consecutively determines the optimal power plant operation for three consecutive days. Moreover, the model considers hourly net transfer capacities to limit electricity transmission across countries and a formulation for medium- and long-term energy storage. We describe these steps in detail in Section\ref{sec:Methodology}. 

Third, we demonstrate the value of sequentially and continuously improving the quality of input variables in fundamental energy system models in the empirical part of the paper. We consider TSO day-ahead load forecasts provided by one of the most used data sources \cite{EntsoeTPe} and day-ahead prices forecasted with the energy system model for Germany, one of the largest and most liquid electricity markets in the world. By capturing and reflecting systematic biases and autoregressive structures, we reduce the mean squared error by 26~\% compared to the TSO-based load forecast. Therefore, market participants' expectations of the day-ahead market clearing can be better reflected. As a result, the mean squared error of the \emph{em.power dispatch} model's price forecast is reduced by nearly 15~\% in hours with high prices using the improved load forecast compared to using the TSO load forecast. By demonstrating that energy system models with the improved load data perform significantly better compared to the TSO data, we provide valuable insights for many stakeholders in the power sector, particularly energy system model developers seeking to improve the validity of their models. Based on these results, we encourage energy system modellers and all users of fundamental input data to be aware of the predictable structure of their errors. In particular, stochastic modelling of the errors significantly reduces the forecast error of input data. It thus improves the quality of input data as part of sequential data preprocessing in real-time and offers the possibility to enhance the output of fundamental energy system models. 

The remainder of the paper is organised as follows. First, we examine the literature on energy systems modelling, data quality and time series modelling in Section\ref{sec:literature}. Section \ref{sec:data} presents the data used in this application. In Section\ref{sec:Methodology}, we provide and explain the methodology for the model improving the load forecasts and the energy system model used to evaluate the impact of the improved load data. The results are presented in Section\ref{sec:results}. Finally, a conclusion is drawn in Section\ref{sec:conclusion}.

\section{Literature}
\label{sec:literature}

Energy system models are widely used in academia, policy-making and industry. Typically, they determine market equilibria, minimising production costs or maximising social welfare. A market's supply and demand sides are equally essential to derive equilibria. Various models have been developed on the demand side using time series of load data as an essential input. On the supply side, models focus on power plants (electricity system models) or gas production (gas systems). Transmission and distribution infrastructure, i.e., connecting supply and demand, can also be included and analysed with energy system models. A strength of these models is that they can provide valuable insights into both causes and effects of current and planned developments, as well as into "what-if" types of analyses. Thus, energy system models are among the most essential methodologies for a successful energy transition.   

Out of a wide range of applications, examples include the determination and assessment of long-term investment decisions for generation and storage capacities (e.g., \cite{NAHMMACHER2016455,SCHILL2018156}) or implications on short-term operational decisions (e.g., \cite{Schill2017}), transmission expansion planning (e.g., \cite{EGERER2021696}, \cite{Sauma2006}), the evaluation of carbon reduction paths (e.g., \cite{VAILLANCOURT2017774}) and support schemes for renewable energy systems (e.g., \cite{KITZING201783})
and the evaluation of interdependencies between energy sectors (e.g., \cite{LIENERT201299} for electricity and gas markets, \cite{HEINISCH2021116640} transport, electricity and district heating, \cite{KOIRALA2021116713} electricity, hydrogen and methane). Moreover, scholars developed stochastic models to assess the impact of uncertainty on a power system \cite{RIEPIN2021116363}, for example, to quantify the expected costs of ignoring uncertainty of critical parameters in the electricity and gas sector. \cite{MOST2010543} provide an overview and classification of stochastic models dealing with uncertainty in the power sector. With regard to uncertainty, scholars analyse the effect of risk preferences as well (e.g., \cite{Moebius2021}, \cite{AMBROSIUS2022105701}).  

In this paper, we focus on the impact of better load forecasts on the estimation of wholesale electricity prices with fundamental market models. Estimating wholesale electricity prices is essential for making optimal economic decisions (e.g., investment and dispatch of various technologies) and policy decisions (e.g., calculating the implications of a coal phase-out). Wholesale electricity prices can be forecasted with multiple methodologies, all with their unique advantages and disadvantages. Energy system models have advantages, e.g., they perform exceptionally well at structural breaks, are based on a broad economic theoretical foundation explaining causality, and provide additional information beyond the forecast. Consequently, much attention has been paid in the literature to the simulation or prediction of electricity prices in energy system models.  
\cite{HIRTH2013218} simulate electricity prices to quantify the drop in the market value of variable renewables. Additionally, \cite{EISING2020104638} quantify market values for renewables generating electricity prices in a future power system with the help of an energy system model.
\cite{Engelhorn2022} derive market prices assuming different weather years and quantify weather-specific market values for a comprehensive database of onshore wind capacities in Germany. \cite{Qussous2022} use an agent-based model with rule-based bidding strategies to reproduce spot prices for the German bidding zone.

Market power and strategic behaviour are other applications of wholesale price forecasts with energy system models. When modelling competitive market prices and comparing them with actual prices, they were able to point to serious problems (e.g., \cite{Muesgens:2006} and \cite{WEIGT20084227} for Germany, and \cite{Borenstein2002} for the United States).

These and many other model applications have a dedicated empirical focus. Thus, the high quality of input data is vital. Concerning load data, a comprehensive literature review of various methods and models for energy demand forecasting is given by \cite{SuganthiLoadReview2012,6508132}. Among others, approaches for standalone load forecasting models are presented by \cite{WeronLoad2005, WeronLoadbook2006, MLLoad2019, enLoad11071899, Rodrigues2018, WU2019, Tan2010, CHEN1995, ALHAMADI2004} and \cite{Yang2013}. For the European electricity sector, data is conveniently gathered and made publicly available by transmission system operators (TSOs) via the transparency platform of the ENTSO-E. The platform is a very ambitious and unique project to provide an extensive data set for electricity markets and is thus both well-known and widely used. However, it is not without its shortcomings (see \cite{HIRTH2018}). For example, \cite{Maciejowska2021} analyse of the quality of load data for the Germany-Luxembourg bidding zone. They detect a bias in TSO load forecasts and develop an alternative load prediction model that incorporates information from these forecasts to remove the bias and thus achieve an enhanced load prediction. \cite{CanceloLoad2008588} analyse the forecast errors of the Spanish day-ahead TSO load forecasts in detail for serial structures and influences of special days such as Christmas holidays or New Year's Eve. 

Given a series of load forecasts with forecasting errors that still show a predictable structure, the method proposed in this paper offers a possibility to enhance such forecasts. We improve the forecasts by modelling and removing predictable parts of the errors. Implicitly, \cite{Yang2013} use a similar step since they remove a structure from their forecasting model (first stage) in a second stage by a time series approach. However, they rely on neural networks, while we propose a very simple time series model. 

In energy system modelling, such steps can be described as data preprocessing or, more precisely, continuous data processing and enhancement with subsequent use. Such continuous data processing is typically not performed for energy system models. We believe this is a methodological gap in the literature and aim to bridge it by providing an approach to sequentially improving input data and sequentially using these continuously improved datasets in an energy system model. We demonstrate the effectiveness in an empirical application, focusing on the effect of better load forecasts for electricity price forecasts derived from energy system models.

\section{Data}
\label{sec:data}

Energy system models require extensive input data to model market equilibria on both the demand and the supply sides. 
Since this paper focuses on a day-ahead time horizon, TSO-based load forecasts published by ENTSO-E may be used as predictors for the demand side. 
However, as was pointed out in the literature section, the quality of these load forecasts is debated and will be improved in this paper. 
In Section\ref{subsec:TSO-based_Load_Data}, we first provide a detailed overview of the TSO-based load forecast data and forecast errors. Moving to the supply side of the energy system model, data on techno-economic parameters for conventional generation, renewables, storage and electricity transmission are of the utmost importance and are presented in Section\ref{subsec:data_dispatch_model}. 

\subsection{TSO-based Load Forecast Data}
\label{subsec:TSO-based_Load_Data}

\begin{figure}[htb]
    \centering
    \includegraphics[width=1\linewidth]{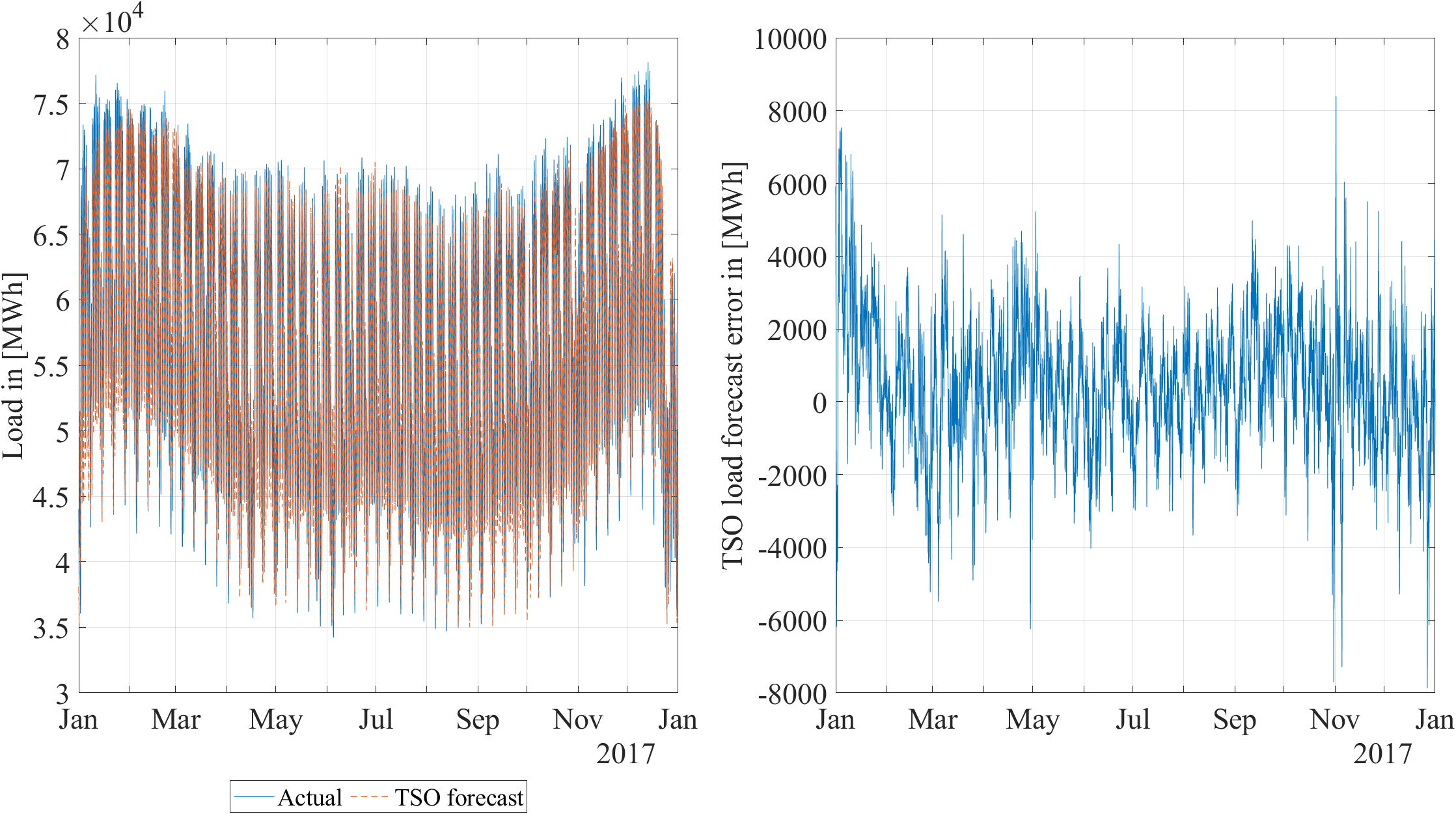}
    \caption{Actual load and TSO's day-ahead load forecast in 2017 (left) and error of TSO's day-ahead load forecast in 2017 (right).}
    \label{im_load2017}
\end{figure}

The load data set we use for our analysis contains hourly day-ahead load forecast data and hourly actual load data from January 1st, 2016, until December 31st, 2019, for Germany and Luxembourg. It was downloaded from \cite{EntsoeTPe} in MWh. Missing values were replaced by the average of the value of the previous week and the week after\footnote{There are 1,105 missing values in the hourly TSO load forecast and 38 missing values in the hourly actual load data.}. An illustration of the time series of the actual load, TSO load forecast and the resulting error, computed as the difference between actual load and load forecast, is shown in Figure \ref{im_load2017}. 

For the considered years, Table \ref{table_descriptives_load_allyears} contains descriptive statistics of the TSO load forecast errors defined as $\epsilon_t \coloneqq L_t - \hat{L}_t$, meaning actual load minus TSO load forecast. Thus, a positive error states an underprediction of load. 

\begin{table}[htb]
	\begin{center}
	\small
		\begin{tabular}{|l||r|r|r|r|r|r|r|r|}
			\hline
				    &	all	&	2016	&	2017	&	2018	&	2019	\\
            \hline
            Mean	&	881.29	&	1,555.38	&	446.50	&	298.60	&	1,222.84	\\
            Median	&	892.75	&	1,468.00	&	479.50	&	395.63	&	1,195.75	\\
            Minimum	&	-20,358.00	&	-7,752.50	&	-7,868.50	&	-20,358.00	&	-7,415.00	\\
            Maximum	&	12,930.75	&	12,930.75	&	8,392.50	&	9,045.25	&	9,635.25	\\
            5\%-quantile	&	-2,477.50	&	-1,684.03	&	-2,413.75	&	-3,180.63	&	-2,266.13	\\
            95\%-quantile	&	4,294.40	&	4,894.20	&	3,099.38	&	3,644.75	&	4,811.38	\\
            Std.	&	2,149.75	&	2,079.19	&	1,746.54	&	2,341.68	&	2,128.55	\\
            LB hypothesis	&	1.00	&	1.00	&	1.00	&	1.00	&	1.00	\\
			\hline
			
		\end{tabular}
		\bigskip
		\caption{Descriptive statistics of TSO load forecast errors for the years 2016 to 2019. Except for LB hypothesis, all variables are given in [$MWh$].}
		
		\label{table_descriptives_load_allyears}
	\end{center}
\end{table}

The TSO forecast data is mean-biased, as discussed in \cite{Maciejowska2021}. In our analysis, we find systematic underpredictions with a mean error of 881.3~MWh across all years and positive mean errors for every year.

\begin{figure}[H]
    \centering
    \includegraphics[width=1\linewidth]{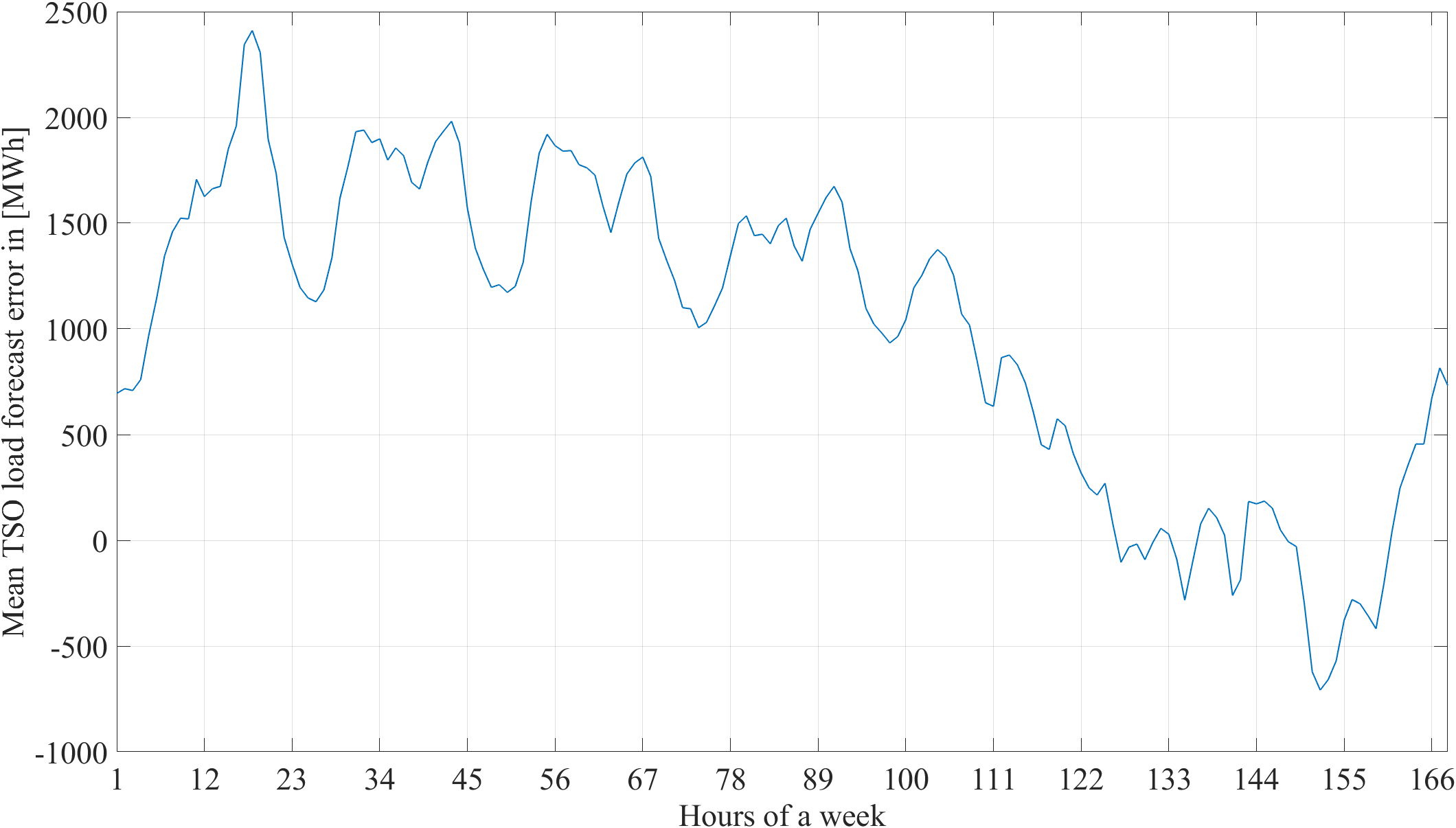}
    \caption{Average weekly pattern of TSO day-ahead load forecast errors from 2016 to 2019.}
     \label{im_TSOerror_weekJul19}
\end{figure}

However, the absolute level of the error and whether the TSO under- or overpredicts in its forecasts depend on the day of the week and the hour of the day. Figure \ref{im_TSOerror_weekJul19} states the averaged hourly forecast errors in a week. Broadly, we can observe underprediction during weekdays and overprediction on the weekends, especially on Saturdays. During the day, in the morning and the evening hours, the error of the TSO day-ahead load forecast is generally positive and higher than in the other hours of the day. With an average error of 943.53~MWh at 6 a.m. and 1,180.48~MWh at 7 p.m., the prediction error in these hours is higher by 7~\% (34~\%) than the mean error of the entire time period considered (compare with Table \ref{table_descriptives_load_allyears}).
These are the hours when the workday begins or ends and where production ramps up or down. Although the standard deviation of the forecast in these hours is not significantly larger than in the other hours, it appears that the load in these hours is still more challenging to forecast on average than in the other hours of the day (see weekday-wise descriptive measurements in Table \ref{table_descriptives_load_allyears_hourly} for more details). 

\begin{figure}[H]
    \centering
    \includegraphics[width=1\linewidth]{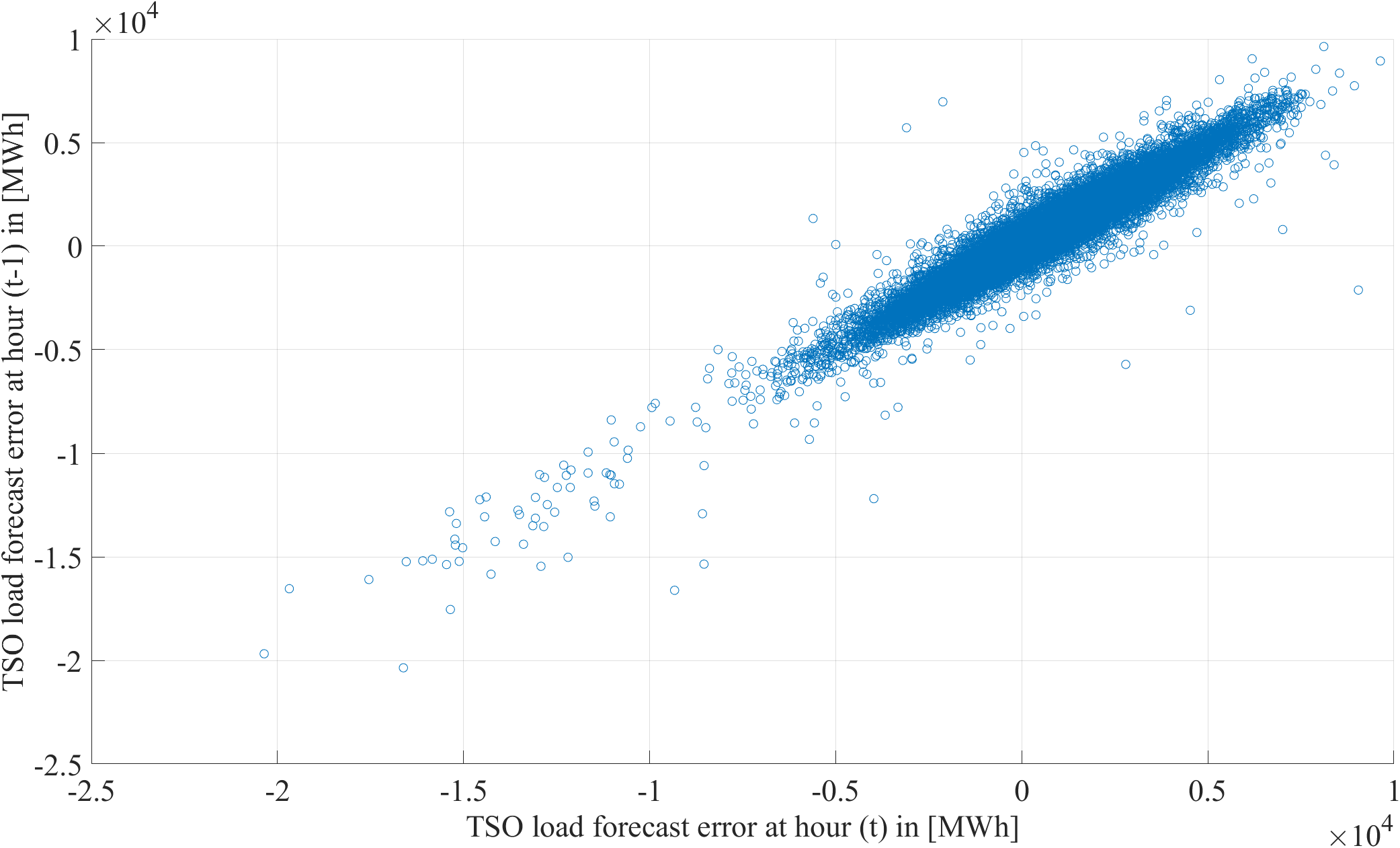}
    \caption{Scatterplot of TSO's day-ahead load forecast error and TSO's day-ahead load forecast error one hour before.}
    \label{im_scatter_TSOerror_t_t-1}
\end{figure}

Finally, we perform Ljung-Box (LB) tests to verify the auto-correlation of the TSO load prediction errors. The null hypothesis at a 5~\% significance level is rejected for all years, which indicates a strong auto-correlation of the errors. Comparing the errors with those one hour before (see Figure \ref{im_scatter_TSOerror_t_t-1}), we can see a highly linear dependence. 

In summary, the load data shows high auto-correlated TSO forecast errors, which average 1.56~\% of the total load's mean. The mean absolute error of the TSO load forecast is 1,776~MWh (3.14~\% of the total load's mean). The TSO forecast errors are biased with some seasonal structures in the bias and are highly auto-correlated. Hence, autoregressive type models could improve the TSO load forecast. 

\subsection{Input data for an energy systems model}
\label{subsec:data_dispatch_model}

Aiming to analyse the impact of improved day-ahead load forecasts on the accuracy of electricity price forecasts, which are derived using an electricity system model, we develop and parameterise a European electricity market model with data from January 1st, 2017, until December 31st, 2019. A meaningful empirical parameterisation of such models requires extensive input data derived from various sources. To model the demand side, the load data presented in the previous Section \ref{subsec:TSO-based_Load_Data} is essential. Furthermore, there is typically an option to shed load during supply scarcities. In our application, we assume the costs for load shedding to be 3,000~€/MWh.

On the supply side, a number of technologies are available for electricity generation and storage. Our energy system model distinguishes ten conventional thermal generation technologies, which form 30 capacity clusters according to a power plant's commissioning year. We provide each of the capacity clusters with different efficiencies, minimum outputs and efficiency losses in part-load operations, which are derived from \cite{Schroeder2013} and \cite{OPSDa}. The capacity, fuel type, generation technology and commissioning date are derived from \cite{EntsoeTPa} and \cite{OPSDa} and \cite{EBC2021}. For power plants on the German market, we additionally use data from \cite{BNetzA2021} and \cite{UBA2020}.
Fuel costs, costs for CO$_2$ emissions and the power plant efficiency determine the variable generation costs of conventional thermal technologies. For fuel costs, we use daily gas prices that are provided by \cite{EEX2021}, monthly coal prices are taken from \cite{Destatis2020}, and monthly oil prices from \cite{Destatis2020}. Fuel costs for nuclear, lignite and waste are derived from \cite{EntsoS}. These are assumed to be constant over time. Prices for CO$_2$ certificates are implemented as weekly values from \cite{Sandbag2020}.

The process of starting up power plants requires the use of fuel, emits CO\textsubscript{2} and leads to material wear in the plant. Data for start-up times, secondary fuel usage and depreciation are derived from \cite{Schroeder2013}. 

The ability to generate electricity depends not only on the installed capacity but also on the technical availability of the plants. Therefore, we consider all scheduled and non-scheduled power plant outages known before the day-ahead market's closure. Hourly outages are derived from \cite{EntsoeTPb}. 

Since combined heat and power (CHP) plants are used in most electricity markets, electricity and heat supplies are linked. To account for this dependency, we provide these units with a must-run condition that ensures their operation at certain minimum output levels. These output levels are derived in two steps. First, we determine an hourly heat-demand factor consisting of a temperature-dependent (spatial heating) and temperature-independent (warm water and process heat) part. The temperature-dependent heat demand is generated with heating degree days using mean temperature data from \cite{OPSDb}. We derive the temperature-independent heat demand using the hourly and daily consumption patterns from \cite{Hellwig2013}. Second, we use the heat-demand factor to allocate annual electricity generation volumes by CHP plants to single hours. The annual technology-specific electricity generation by CHP units is taken from \cite{EC2021}.

In addition to conventional thermal technologies, we consider renewable energy sources (RES), energy storage, hydro-reservoirs and run-of-river. Intermittent RES such as onshore wind, offshore wind and photovoltaics (PV) are implemented by hourly availability factors that are derived from feed-in forecasts from \cite{EntsoeTPc}. We do not also improve these forecasts by sequentially modelling their forecast errors in order to clearly measure the impact on the quality of the price forecast when we improve the forecast of the variable that not only offers the greatest potential for improvement but is also most strongly correlated with day-ahead electricity spot market prices. Biomass is implemented as base-load as the historic operation is at a constant level (compare \cite{EntsoeTPd}).

We exclusively consider pumped storage plants (PSP) for energy storage that actively charge and discharge. The overall turbine capacity of PSPs is made available by \cite{EntsoeTPa}, and the efficiency of a storage cycle is around 75~\% (\cite{Schroeder2013}). For PSPs, the energy storage capacity and the turbine capacity are linked. Assuming an energy-power factor (epf) of nine, the plant can generate electricity at full load for nine hours until the storage is empty.

Long-term PSP, as well as hydro-reservoirs, are assigned a variable generation cost, i.e., the value for water consumption. Using historical electricity prices from \cite{EntsoeTPg} and the observed generation and pumping activities in the respective hour from \cite{EntsoeTPd}, a step-wise merit-order for long-term PSP and hydro-reservoirs is constructed. Run-of-river and mid-term PSP\footnote{Note that we call it mid-term because we focus on the day-ahead market with an hourly granularity, as opposed to short-term storage with an intra-hourly resolution closer to time of delivery.} are subject to seasonal variations, which we acknowledge by a monthly availability factor derived from historical generation data from \cite{EntsoeTPd}.

The German electricity market is highly integrated into the European system. Total interconnector capacity amounts to 27~GW, which is more than 30~\% of the German peak load.\footnote{Note that the availability of the interconnectors depends on various factors (e.g., congestion within a market zone).} Both annual aggregated exports (around 13~\% of annual German consumption in 2019) and imports (around 7~\% in 2019) are significant. Hence, we parameterise a Pan-European electricity market model which includes the bidding zones of most EU-27 member states \footnote{Bulgaria, Cyprus, Greece, Iceland, Ireland, Malta and Romania are not included.}, Norway, Switzerland and the United Kingdom.\footnote{Note that we aggregate the bidding zones of Spain and Portugal to one market, 'Iberian peninsula', and the bidding zones of Lithuania, Estonia and Latvia to one market, 'Baltic'. Also, note that we consider the different bidding zones within countries. However, we aggregate the following zones: in Norway NO1-NO5, in Sweden SE1-SE3, and in Italy all zones except IT-North.} Within Germany, day-ahead electricity prices are derived following the principle of a bid-based economic dispatch, neglecting the physical transmission constraints within the market zone. Since the energy system model has its focuses on the analyse of day-ahead prices, we follow this approach and treat all of Germany, plus Luxembourg, as one bidding zone.\footnote{Note that the market area, Germany-Luxembourg-Austria, was split into two market zones (Germany-Luxembourg and Austria) in 2018. Our model accounts for this fact.} Thus, in total, we include 23 different markets in the analysis, which will be referred to as 'nodes' in the formal model, connected by net transfer capacities (NTCs). We implement hourly day-ahead forecasts for NTCs that are made available by \cite{EntsoeTPf} and \cite{JAO2021}.

As the data parameterisation may be interesting for numerous stakeholders but is difficult and time-consuming to replicate, we publish our input data in the supplementary material: \\ \hyperlink{https://github.com/ProKoMoProject/Enhancing-Energy-System-Models-Using-Better-Load-Forecasts}{github.com/ProKoMoProject/Enhancing-Energy-System-Models-Using-Better-Load-Forecasts}.

\section{Methodology}
\label{sec:Methodology}
In the following, we present our two components to analyse the value of improved day-ahead load forecasts for electricity price forecasts derived by an electricity system model: a time series model for the sequential load data preprocessing and improvement in Section \ref{ss_Error Correction Model} and the dispatch market model that is used to generate price estimators in Section\ref{subsec:Dispatch Model}.

\subsection{Model for load forecast error}
\label{ss_Error Correction Model}

To improve load forecasts, we use a well-known time series approach that achieves a trade-off between performance and complexity. The approach is based on the idea of forecasting the TSO load forecast error and using this to enhance the load prediction. Thus, we model the time series of forecast errors. For this reason, and to obtain a low-parameter model, we do not use exogenous variables such as feed-in of renewable energy or weather in our model for forecasting the load forecast error, in contrast to the main load forecasting methods in the literature, which include temperature and weather data in particular, e.g., \cite{CanceloLoad2008588, ALHAMADI2004, Amjady2001, WU2019, 9311801}. We propose a purely endogenous time series approach that can be applied using TSO load forecast error alone as input data. It is detached from the outgoing model, which in general already includes exogenous variables. 
With forecasting the forecast error, the resulting load prediction $\hat{L_t}^*$ at time $t$ is then given by 

\begin{equation}
    \begin{aligned}
        \hat{L_t}^* = \hat{L_t} + \hat{\epsilon_t},
    \end{aligned}
\end{equation}

where $\hat{L_t}$ is the original TSO load prediction and $\hat{\epsilon_t}$ is our forecasted TSO load prediction error. Thus, $\hat{L}^*$ is an improved load forecast in which we adjust the original forecast for predictable structure in its error.

For the overall setup, the subindex $t$ will denote consecutive hours. So, $\hat{L}_1$, for instance, is the load forecast for the first hour of the considered time period and $\hat{L}_{123}$ is the forecast for the hour $123$. This fits best into the observation process of the actual load data. For example, in contrast to electricity prices, for which we observe a realisation of 24 daily hourly prices at the same time, load data can theoretically be observed hour by hour. For day-ahead electricity prices, alternative parameterisations such as modelling every day as a 24-dimensional vector, or using 24 time series each for one hour of the day, would be more appropriate (see, e.g., \cite{Ziel2018}).

Furthermore, we decompose the time series into the sum of a seasonal component and a remaining stochastic component. As we do not observe any trend in the forecast error data in Section\ref{subsec:TSO-based_Load_Data}, we do not use the usual trend component of such decomposition models (see, e.g.,  \cite{Luetkepohl2005, Hyndman21, BoxJenkins2016} for comprehensive introductions into time series models). Together, the model is  

\begin{equation}
\label{eqn_ecm_formula}
\begin{aligned}
    \epsilon_t = SC_t + RC_t, \\
\end{aligned}
\end{equation}

where $\epsilon_t$ is the TSO load forecast error, $SC_t$ is a seasonal and $RC_t$ is the remaining component at time $t$. 

The forecast errors' average sizes depend on the specific hour of the week (see Section\ref{subsec:TSO-based_Load_Data}), so the seasonal component $SC_t$ captures a weekly season, consisting of an average value for each of the 24x7 hours of a week. This means addressing the hour of the day and the day of the week with a total of 168 dummy variables, as given by 
\begin{center}
    \small \[ 
 HoW_t^{h,d} = 
  \begin{cases} 
  \text{1,} & \text{if $t$ is the h-th hour of the d-th day of the week,}\\ 
  \text{0,} & \text{otherwise.}
  \end{cases}
\] \end{center}
Here $h = 1, ..., 24$ denote the hours of a day and $d = 1$ (Monday), \ldots, $7$ (Sunday) the weekdays of a week.

The seasonal component $SC_t$ for time $t$ is now defined by Eq. \ref{eqn_Sc_t} with \ref{eqn_HS} being the average of TSO forecast errors from the hours of a week from the time period used to estimate the model (e.g., the last $l_w$ hours).
\begin{equation}
    \label{eqn_Sc_t}
\begin{aligned}
SC_t = \sum_{h=1}^{24} \sum_{d=1}^{7} HoW_t^{h,d} \cdot HS^{h,d}, 
\end{aligned}
\end{equation}
%with 
\begin{equation}
    \label{eqn_HS}
    \begin{aligned}
        HS^{h,d} \coloneqq \frac{\sum_{s=t-1}^{t-l_w-1} \epsilon_s \cdot HoW_s^{h,d}}{\sum_{s=t-1}^{t-l_w-1} HoW_s^{h,d}}, \\
    \end{aligned}
\end{equation}

The rest of the time series $RC_t = \epsilon_t - SC_t$ is modelled by the econometric SARMA $(1,1)$x$(1,1)_{24}$ model given in Eq. \ref{eqn_Sarma}, i.e., a (S)easonal (A)uto(R)egressive (M)oving (A)verage model. Here, the value $RC_t$ at hour $t$ depends on its previous value at $t-1$ as well as the previous model error $\psi_{t-1}.$ Additionally, the model contains a 24-hour seasonal part which captures stochastic seasonal behaviour in contrast to the more deterministic seasonal structure filtered by $SC_t.$ Formally, the seasonal part leads to direct effects of all variables lagged by another 24 hours on $RC_t$ as given in detail in Eq. (\ref{eqn_Sarma}).

\begin{equation}
    \label{eqn_Sarma}
    \begin{aligned}
        RC_t =& \phi_0 + \phi_1 \cdot RC_{t-1} + \phi_{24} \cdot RC_{t-24} - \phi_1 \phi_{24} \cdot RC_{t-25} \\
    & + \omega_1 \cdot \psi_{t-1} + \omega_{24} \cdot \psi_{t-24} + \omega_1 \omega_{24} \cdot \psi_{t-25}\\
    & + \psi_t,
    \end{aligned}
\end{equation}
where the innovations are assumed to be homoscedastic and normally distributed, which means $\psi_t \sim N(0,\sigma^2_\epsilon)$. Assuming a normal distribution for the innovations is a simplification and idealisation. 
 
We calibrate and estimate the model on a rolling window. The window length, denoted by $l_w$, is an integer multiple of 24 and thus contains full days only. The window is also rolled over full days in each step to further reflect the daily availability of load data and thus the error of the TSO's load forecast. In this work, we decide on one window length $l_w$ to estimate the model. Alternatively, one could average multiple models calibrated on different window lengths, e.g., as proposed in \cite{Maciejowska2021, Ziel2018, Marcjasz2018}. However, in this paper, where the simplicity and usability of the model are important considerations, we believe such an increase in complexity would not be justified. 

The estimated model is used to recursively (i.e., on an hour-by-hour basis) predict the hours of the next day. Since we rely on an autoregressive time series model, we need load data from the last hours for prediction, which enter the model as explanatory variables. Although load generation can theoretically be observed hourly, in practice, the load values of the previous hours are available with a time lag, meaning they may not be available as explanatory variables when forecasting the following hours. A solution is to replace unavailable variables with recursively forecasted variables based on the last available observations.

To ensure data availability in the sense of a day-ahead forecast at all times, we only use load observations up to yesterday's last hour for TSO data as inputs if we make predictions today for tomorrow. Today's hours must be replaced by forecasts based on yesterday. More clearly, let $t=8785$ be the first hour of January 1st, 2017, for simplicity and let $x$ be the hour of January 1st from which we forecast the next day's hours. In the further course, we assume $x=12$, so we forecast the next day's hours between 11:00 and 12:00 a.m. today.
Depending on availability, real TSO load forecast errors $\epsilon_t$ enter our model or forecasted ones. For hour $t \leq x-12$, we use the observed real errors ${\epsilon}_t$ and the forecasted ones $\hat{\epsilon}_t$ for $t > x-12$. We want to predict the load for the next day's 24 hours, thus, $x+13$ to $x+37$. Due to the information delay and ensuring data availability, we do not indicate the actual load of hours $x-11$ to $x-1$. We also have no information about the hours $x$ to $x+12$ lying in the future. 
For this reason, we first estimate the model based on the last available $l_w$ observations (i.e., of hours $x-12-(l_w+1)$ to $x-12$. From that, we predict the errors of the TSO load forecast of the next $48$ hours $x-11$ to $x+37$, i.e., of the hours of January 1st and 2nd, and use the last $24$ predicted values. Thus, at hours $x+13$ to $x+37,$ for improving the original load forecasts of the following day. Note that by rolling over the estimation window daily, we ensure that the prediction of TSO forecast errors for all load periods of one day is based on the same estimated model. 

The proposed model is implemented in MATLAB\textregistered. The code, used data and the generated result are provided on GitHub: \\
\hyperlink{https://github.com/ProKoMoProject/Enhancing-Energy-System-Models-Using-Better-Load-Forecasts}{github.com/ProKoMoProject/Enhancing-Energy-System-Models-Using-Better-Load-Forecasts}.

\subsection{Energy System Model}
\label{subsec:Dispatch Model}

We develop a new energy system model, the \emph{em.power dispatch} model, to derive wholesale day-ahead price forecasts. The model is formulated as a linear optimisation problem minimising total system costs and includes a detailed representation of central techno-economic aspects of the European electricity sector. In particular, the model dispatches various generation technologies to satisfy electricity demand. In addition to power plant dispatch in Germany, the model considers international trade between the markets described in \ref{subsec:data_dispatch_model}, electricity production by combined heat and power plants, energy storage and control power provision. To ensure a linear formulation of such a highly complex system, we form capacity clusters, parameterised as described in \ref{subsec:data_dispatch_model}. Within each technology cluster, capacity can be started-up and electricity can be produced in marginal increments (see, e.g., \cite{Muesgens:2006}). The advantage of this approach is twofold. First, computational efforts are reduced. Second, the marginal of the demand restriction is differentiable at each point and can thus be interpreted as a wholesale market price estimator. Additionally, the accuracy of modelling large energy systems, in particular, remains reasonably high (see \cite{MuesgensandNeuhoff:2006}).

Considering all economic and technical restrictions, the model solves the cost minimisation problem and determines i) the optimal dispatch decision for all considered infrastructure elements, such as generation technologies, energy storage and cross-border transmission capacities, and ii) the short-run marginal system cost that determines the price estimator for the day-ahead market in hourly resolution.

\begin{figure}[H]
    \centering
    \includegraphics[width=0.6\linewidth]{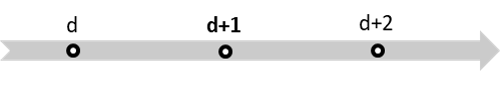}
    \caption{Illustration of the rolling window.}
    \label{figure_rolling_window}
\end{figure}

Furthermore, as our research analyses the impacts on day-ahead price forecasts, we set up the model to reflect the information available to market participants on the day before delivery. We thus consider that market participants do not have perfect foresight for the upcoming days. We achieve this with a rolling window model that is repeatedly solved and provides information for 24 day-ahead hours of one "target day" in each model run. To reduce the problem of starting and ending values, in particular for power plant start-ups and pump storage plants, each model run includes three days, as shown in \ref{figure_rolling_window}. In this setting, the 24 hours of the respective target day are represented by the second day of the horizon (d+1). This is following the EPEX spot market organisation, where 24 hourly day-ahead prices are determined at 12 p.m. on the day before delivery (d). In addition to the target day d+1, we also include the day before (d) and the day after (d+2). Note that we include a water value to increase the accuracy of seasonal hydro-storage modelling.

\FloatBarrier 

As with the improvement of the load forecast, this approach is repeated continuously ("rolling window"), once for each day of the observation period. At each iteration, the input data for d+1 and d+2 are limited to the values available on day d (i.e. forecasts), so that the incoming day-ahead load forecast is successively improved and processed in our approach. Correctly parameterised, our model uses the same data as market participants (e.g., energy suppliers, direct marketers, investment banks) when forecasting the day-ahead prices to optimise their portfolio. Given this day-ahead focus of our analysis, installed and available capacities are exogenous. The model endogenously optimises power plant dispatch only. 

Our rolling window approach to forecasting hourly prices implies that we forecast three years with 365 daily model runs each year. As each model run comprises 72 hourly dispatch decisions with numerous variables in 23 model regions, the total number of variables is 340 million. In the following, we present the mathematical formulation of our model. The model is coded in GAMS\footnote{GAMS General Algebraic modelling System (https://www.gams.com/)}. The entire code is provided on GitHub: \hyperlink{https://github.com/ProKoMoProject/Enhancing-Energy-System-Models-Using-Better-Load-Forecasts}{github.com/ProKoMoProject/Enhancing-Energy-System-Models-Using-Better-Load-Forecasts}. A nomenclature containing all indices, parameters and variables of the energy system model formulation is provided in \ref{Nomenclature}.

The objective function in Eq. \ref{eqn:TotalCost} minimises total system costs and accounts for all costs that generation units face in the short-term. We include costs at full load operation ($vc_{i,n,t}^{FL}$), additional costs for units that operate at partial load ($vc_{i,n,t}^{ML}-vc_{i,n,t}^{FL}$), and start-up costs ($sc_{i,n,t}$). Note that we apply a linear formulation of the unit commitment, and all units have to produce at least a minimum output level. Additionally, we account for load shedding costs ($voll$) and penalty payments for curtailing renewables ($curtc$).

Since we apply our model with a rolling window, we consider three days in each model run. Modelling an additional day before and after the target day seems appropriate for storages with large energy-to-power ratios, which are essentially operated on a daily cycle (e.g., the largest German pump storage facility, Goldisthal, can store enough energy for nine hours of full load operation). However, other storages (both PSP and seasonal storages without pumps) have a storage cycle longer than three days. Therefore, we model two types of PSP, first as mid-term storage that operates a storage cycle within a three-day horizon, and second as long-term storage that operates a storage cycle longer than three days. The dispatch of mid-term storage is determined endogenously, with the exogenous restriction that they both start and end the cycle with reservoir levels at 30~\%. The approach is different for long-term PSPs, which are assigned a water value ($wv_{stl,n,t}$) that is implemented as a variable cost factor for electricity generation ($G_{stl,n,t}$) and consumption ($CL_{stl,n,t}$). We assume that 70~\% of the pump storage capacity is optimised in the medium-term. The remaining 30~\% are long-term PSPs. 

Compared to pumped storage plants, hydro-reservoirs have a natural water feed-in and do not perform a pumping process. However, the water budget for electricity generation is limited according to seasonal inflow volumes. Therefore, we also apply a water value for electricity generation by hydro-reservoirs.

\begin{equation}
\label{eqn:TotalCost}
\begin{aligned}
    \displaystyle\min TC= & \sum_{i,n,t} G_{i,n,t}\cdot vc_{i,n,t}^{FL} + \sum_{i,n,t} SU_{i,n,t}\cdot sc_{i,n,t} \\
    &+ \sum_{i,n,t} (P_{i,n,t}^{on} - G_{i,n,t})\cdot (vc_{i,n,t}^{ML}-vc_{i,n,t}^{FL})\cdot g_i^{min} / (1-g_i^{min})  \\
   & - \sum_{stl,n,t} CL_{i,n,t}\cdot wv_{stl,n,t} + \sum_{n,t} SHED_{n,t}\cdot voll  \\
   & + \sum_{n,t} CURT_{res,n,t}\cdot curtc 
\end{aligned}
\end{equation}

Market clearing is ensured by Eq. \ref{eqn:EnergyBalance}: for all T hours of the given rolling window, demand ($d_{n,t}$) must equal the sum of generation ($G_{i,n,t}$), load shedding ($SHED_{n,t}$) and electricity imports ($FLOW_{nn,n,t}$), reduced by electricity consumption of mid-term energy storage ($CM_{stm,n,t}$) and long-term energy storage ($CL_{stl,n,t}$) and electricity exports ($FLOW_{n,nn,t}$). 

\begin{equation}
\label{eqn:EnergyBalance}
\begin{aligned}
     d_{n,t}= & \sum_i G_{i,n,t} - \sum_{stm\subset I} CM_{stm,n,t} -\sum_{stl\subset I} CL_{stl,n,t} + SHED_{n,t} \\
    & + \sum_{nn} (FLOW_{nn,n,t} - FLOW_{n,nn,t})  \\
    &\forall n,nn\in N, t\in T \\ 
\end{aligned}
\end{equation}

The dual variable of the demand constraint Eq. \ref{eqn:EnergyBalance} is used as an hourly day-ahead wholesale electricity price estimator. As we want to analyse how well these price estimators based on different demand forecasts fit real-world day-ahead prices, we compare them and compute error measures.

Electricity generation by capacity cluster is limited by an upper and a lower bound. The upper bound is formalised in Eq. \ref{eqn:GenMax} and ensures that electricity generation does not exceed the running capacity ($P_{i,n,t}^{on}$) in the cluster. The possible electricity generation by running capacity is further limited by the reserve for positive control power provision ($PCR_{i,n,bp},SCR_{i,n,bs}^{pos}$). The lower bound is presented in Eq. \ref{eqn:GenMin} and states that running capacities must operate at least at a minimum power level, including the capacity reserved for negative control power provision ($PCR_{i,n,bp},SCR_{i,n,bs}^{neg}$). Note that primary control power ($PCR_{i,n,bp}$) in Germany is provided synchronously, i.e., a unit has to provide both positive and negative primary control power. Different products for positive and negative control power were introduced for secondary control power. Since fast-reacting units (e.g., hydro- and open-cycle gas turbines) can be started-up to provide a positive-minute reserve, the effect on the running capacities is neglected. In addition, we assume that a negative-minute reserve is provided by multiple market players, not necessarily by power plants. The hours that belong to bidding blocks are mapped for primary control power by $bp$ and secondary control power by $bs$.

\begin{equation}
\label{eqn:GenMax}
\begin{aligned}
    \displaystyle G_{i,n,t}  \leq  P_{i,n,t}^{ON} - PCR_{i,n,bp|t\in bp} - SCR_{i,n,bs|t\in bs}^{pos} \\
    \forall bp\in BP, bs\in BS, i\in I, n\in N, t\in T \\ 
\end{aligned}
\end{equation}

\begin{equation}
\label{eqn:GenMin}
\begin{aligned}
    \displaystyle P_{i,n,t}^{on}\cdot g_i^{min} + PCR_{i,n,bp|t\in bp} + SCR_{i,n,bs|t\in bs}^{neg}  \leq   G_{i,n,t}  \\
    \forall bp\in BP, bs\in BS, i\in I, n\in N, t\in T \\ 
\end{aligned}
\end{equation}

The running capacity of a power system is limited by the installed capacity ($cap_{i,n,t}$) in combination with either the availability factor ($af_{i,n,t}$) or power plant outages ($out_{i,n,t}$), as shown in Eq. \ref{eqn:PonMax}. For thermal generation capacities, we use hourly power plant outages. Renewables are provided with an hourly availability factor and hydroelectric units with a monthly availability factor.

\begin{equation}
\label{eqn:PonMax}
\begin{aligned}
    \displaystyle P_{i,n,t}^{on} \leq  cap_{i,n,t} \cdot af_{i,n,t} -out_{i,n,t} \hspace{1cm}  \forall i\in I, n\in N, t\in T \\ 
\end{aligned}
\end{equation}

Eq.\ref{eqn:startup} tracks start-up activities ($SU_{i,n,t}$) that increase the running capacity from one hour to another. Due to the non-negativity condition, start-ups are either positive or zero. 

\begin{equation}
\label{eqn:startup}
\begin{aligned}
    \displaystyle P_{i,n,t}^{on} - P_{i,n,t-1}^{on} \leq  SU_{i,n,t}  \hspace{1cm}  \forall i\in I, n\in N, t\in T \\ 
\end{aligned}
\end{equation}

The delta between available feed-in from intermittent renewables and their actual generation defines the curtailment of renewables ($CURT_{res,n,t}$), as shown in Eq. \ref{eqn:RES}. 

\begin{equation}
\label{eqn:RES}
\begin{aligned}
    \displaystyle cap_{res,n,t}\cdot af_{res,n,t} = G_{res,n,t} + CURT_{res,n,t} \hspace{1cm}      \forall res\in I, n\in N, t\in T \\ 
\end{aligned}
\end{equation}

Some power plants are active in the heat market in addition to the electricity market. The model thus implements a must-run condition for such units on the electricity market, which varies over time (e.g., higher in the winter season due to space heating). Depending on hourly heat demand, Eq. \ref{eqn:CHP} states that the output of a combined heat and power unit is at least equal to the electricity generation linked to the heat production ($chp_{i,n,t}$).

\begin{equation}
\label{eqn:CHP}
\begin{aligned}
    \displaystyle chp_{i,n,t} \leq  G_{i,n,t} \hspace{1cm}  \forall i\in I, n\in N, t\in T \\ 
\end{aligned}
\end{equation}

Eq. \ref{eqn:CHP} constraints the cross-border electricity transfer ($FLOW_{n,nn,t}$) by the net transfer capacity ($ntc_{n,nn,t}$).

\begin{equation}
\label{eqn:Trade}
\begin{aligned}
    \displaystyle FLOW_{n,nn,t} \leq  ntc_{n,nn,t} \hspace{1cm}    \forall  n,nn\in N, t\in T \\ 
\end{aligned}
\end{equation}

Eq.\ref{eqn:StorageLevel} describes the state of the storage level of a mid-term storage. The storage level is increased by the generation ($G_{stm,n,t}$) and decreased by the consumption while charging ($ST_{stm,n,t}^{in}$). The efficiency of an entire storage cycle ($\eta_{stm}$) is assigned to the charging process.

\begin{equation}
\label{eqn:StorageLevel}
\begin{aligned}
    \displaystyle &SL_{stm,n,t} =  SL_{stm,n,t-1} - G_{stm,n,t} + CM_{stm,n,t} \cdot \eta_{stm} \\
  &\forall  stm\in I, n\in N, t\in T \\ 
\end{aligned}
\end{equation}

The maximum energy storage capacity ($SL_{stm,n,t}$) of a mid-term storage is defined by the maximum installed turbine capacity times an energy-power factor ($epf$), as shown in Eq. \ref{eqn:maxSL}.

\begin{equation}
\label{eqn:maxSL}
\begin{aligned}
    \displaystyle SL_{stm,n,t} \leq  cap_{stm,n,t} \cdot epf \hspace{1cm}  \forall  stm\in I, n\in N, t\in T \\ 
\end{aligned}
\end{equation}

Eq. \ref{eqn:maxTurbine} restricts the turbine and pumping capacity, where the pumping capacity is assumed to be lower than the turbine capacity.

\begin{equation}
\label{eqn:maxTurbine}
\begin{aligned}
    \displaystyle G_{stm,n,t} + 1.1\cdot CM_{stm,n,t} \leq  cap_{stm,n,t}  \hspace{1cm} \forall  stm\in I, n\in N, t\in T \\ 
\end{aligned}
\end{equation}

At the beginning and end of each model run, all mid-term storages must be filled with 30~\% of their energy level (Eq. \ref{eqn:SLstart} and \ref{eqn:SLend}).

\begin{equation}
\label{eqn:SLstart}
\begin{aligned}
    \displaystyle SL_{stm,n,tfirst} =  0.3\cdot cap_{stm,n,t} \hspace{1cm} \forall stm\in I, n\in N, tfirst\in T  \\ 
\end{aligned}
\end{equation}

\begin{equation}
\label{eqn:SLend}
\begin{aligned}
    \displaystyle SL_{stm,n,tlast} =  0.3\cdot cap_{stm,n,t}  \hspace{1cm}   \forall  stm\in I, n\in N, tlast\in T  \\ 
\end{aligned}
\end{equation}

Long-term storage is not subject to a storage mechanism. However, the electricity generation and consumption of long-term storage units are also restricted by the installed capacity of long-term storage by Eq. \ref{eqn:MaxStoreLong}.

\begin{equation}
\label{eqn:MaxStoreLong}
\begin{aligned}
    \displaystyle G_{stl,n,t} +  CL_{stl,n,t} \leq  cap_{stl,n,t} \hspace{1cm}   \forall stl\in I, n\in N, t\in T \\ 
\end{aligned}
\end{equation}

Eqs. \ref{eqn:CP_primary}, \ref{eqn:CP_sec_pos} and \ref{eqn:CP_sec_neg} ensure the control power provision for primary, positive secondary and negative secondary control power. 

\begin{equation}
\label{eqn:CP_primary}
\begin{aligned}
    \displaystyle \sum_i PCR_{i,n,bp} = pr_{n} \hspace{1cm} \forall  bp\in BP, n\in N \\ 
\end{aligned}
\end{equation}

\begin{equation}
\label{eqn:CP_sec_pos}
\begin{aligned}
    \displaystyle \sum_i SCR_{i,n,bs}^{pos} = sr_{n}^{pos} \hspace{1cm}  \forall  bs\in BS, n\in N \\ 
\end{aligned}
\end{equation}

\begin{equation}
\label{eqn:CP_sec_neg}
\begin{aligned}
    \displaystyle \sum_i SCR_{i,n,bs}^{neg} = sr_{n}^{neg} \hspace{1cm} \forall  bs\in BS, n\in N \\ 
\end{aligned}
\end{equation}

The non-negativity constraint is presented in Eq. \ref{eqn:nonnegative}.

\begin{equation}
\label{eqn:nonnegative}
\begin{aligned}
    0  \leq   & CL_{stl,n,t}, CM_{stm,n,t}, CURT_{res,n,t}, G_{i,n,t}, FLOW_{n,nn,t},\\
    & P_{i,n,t}^{on}, PCR_{i,n,bp}, SCR_{i,n,bs}^{neg}, SCR_{i,n,bs}^{pos}, SHED_{n,t}, SL_{stm,n,t}, SU_{i,n,t}  \\
    & \forall  bp\in BS, bs\in BS, i\in I, n\in N, t\in T \\ 
\end{aligned}
\end{equation}

We use both models presented alternately. To predict the next day, we first forecast the load forecast error with the load forecast improvement model and thus enhance the day-ahead load forecast. As one input data, it enters the power system model, which estimates the next day's prices using the presented approach. This sequence is repeated continuously day by day over the rolling window for all points in time in our observation period.

\section{Results}
\label{sec:results}
Our paper explores two different methodologies that are combined. It presents a forecast error improvement model for load forecasts based on data from ENTSO-E, and it develops the energy system model \emph{em.power dispatch} which is built for day-ahead wholesale price forecasts. We present the results accordingly. First, we show the performance of the model for the load forecast error using statistical data and different error measures for various time periods of the enhanced load forecast. Second, we analyse the impact of the improved forecast on the resulting price estimates of the \emph{em.power dispatch} model. Therefore, we compare the resulting price estimators generated with the original TSO load forecast $\hat{L}$ and the enhanced load forecast $\hat{L}^*$ with the actual price observed at the day-ahead market using several error measures: mean squared error (MSE), root mean squared error (RMSE), and mean average error (MAE). 

\subsection{Improved Load Data and Achieved Error Reduction}
\label{subsec:ImprovedLoadData}
In the following, we quantify the TSO forecast error improvement model described in Section\ref{ss_Error Correction Model}. Therefore, we compare the improved load forecast $\hat{L}^*$ and the TSO load forecast $\hat{L}$ with actual load data $L$. For the error improvement model, we use a rolling window width of one year (i.e., $l_w = 8760$), which yields the lowest (out of sample) error measures compared with a width of three months and six months. For this reason, the prediction of the forecast error, and thus the out-of-sample period, begins on January 1st, 2017. 
Table \ref{table_errormeasures_load_allyears} shows the mean and standard deviation of the TSO load forecast error and the enhanced load forecast error, the error measures MSE, RMSE and MAE of the TSO load forecast and of the enhanced load forecast, as well as the percentage improvement.

\begin{table}[H]
	\begin{center}
	\small
		\begin{tabular}{|l|l||r|r|r|r|r|r|r|}
			\hline
				&	year	&	all	&	2017	&	2018	&	2019	\\
			\hline
			Mean	&	TSO	&	655.98	&	446.50	&	298.60	&	1,222.84	\\
            	&	Impr.	&	-98.89	&	-229.06	&	-36.32	&	-31.29	\\
            \hline
            Std.	&	TSO	&	2,125.72	&	1,746.54	&	2,341.68	&	2,128.55	\\
            	&	Impr.	&	1,743.89	&	1,465.74	&	2,043.67	&	1,665.56	\\
            \hline
            	&	TSO	&	4,948,990.80	&	3,249,416.14	&	5,572,010.71	&	6,025,545.56	\\
            MSE	&	Impr.	&	3,050,928.19	&	2,200,617.23	&	4,177,415.90	&	2,774,751.43	\\
            	&	\% Improvement	&	38.35	&	32.28	&	25.03	&	53.95	\\
            \hline
            	&	TSO	&	2,224.63	&	1,802.61	&	2,360.51	&	2,454.70	\\
            RMSE	&	Impr.	&	1,746.69	&	1,483.45	&	2,043.87	&	1,665.76	\\
            	&	\% Improvement	&	21.48	&	17.71	&	13.41	&	32.14	\\
            \hline
            	&	TSO	&	1,691.37	&	1,396.45	&	1,726.67	&	1,951.00	\\
            MAE	&	Impr.	&	1,253.95	&	1,106.12	&	1,372.55	&	1,283.17	\\
            	&	\% Improvement 	&	25.86	&	20.79	&	20.51	&	34.23	\\
			\hline
			
		\end{tabular}
		\bigskip
		\caption{Means, standard deviations and error measures (MSE, RMSE, MAE) for the original TSO day-ahead load forecast (TSO) and the improved day-ahead load forecast (Impr.). MSE is given in [$MWh^2$], and all other variables in [$MWh$].}
		
		\label{table_errormeasures_load_allyears}
	\end{center}
\end{table}

While the load was severely underestimated in the TSO forecast with a mean of 656.0~MWh, it is slightly overestimated in the improved model with -98.9~MWh. Looking at the individual annual mean values, the high negative value in 2017 is particularly striking. The reason for this is the very strong underestimation of the TSO load forecast in 2016, with an average deviation of 1555.4~MWh (see Section\ref{subsec:TSO-based_Load_Data}). The influence of errors from the year 2016 has a large impact due to the rolling window period of 365 days, especially on the model estimates of the first days and months of 2017. A shorter window period of three months sinks the annual mean value of 2017 but has a minor improvement in error measures (see \ref{table_errormeasures_3mRW_allyears}). 

The standard deviation of the improved load forecast is lower than the standard deviation of the TSO load forecast across all years.

The error measures MSE and MAE given in Table \ref{table_errormeasures_load_allyears} show a significant improvement of the load forecast. With an RMSE of 2,224.6~MWh, we achieve a 21.48~\% improvement over the TSO load forecast for the period from January 1st, 2017, to December 31st, 2019. The most considerable improvement can be observed in 2019 with 32.14~\%. A breakdown of the improvement among the components (seasonal and remaining) of the model shows that both the seasonal and remaining components account for a large share of the improvement, and neither component dominates. 

\cite{Maciejowska2021} also improve the TSO load forecast. From October 1st, 2016, to September 30th, 2019, they achieve an enhancement in RMSE over 365 days from a minimum of 23.71~\% to a maximum of 34.38~\%. Comparing both, achieving a slightly higher improvement also means using a multivariate modelling framework with six different rolling window widths, and consequently six model estimates and six point forecasts for each hour of the forecast period. Our approach is intended to allow a user with less modelling expertise and computational capacity to enhance the commonly used TSO forecast of load. With a less complex, univariate model, we still achieve substantial improvement and thus achieve error measurements that are comparable with error measurements in the literature, e.g., \cite{DO201692, Ziel2018_germanload, 9311801}. 

\begin{figure}[H]
\centering
    \subfigure{\includegraphics[width=0.49\linewidth]{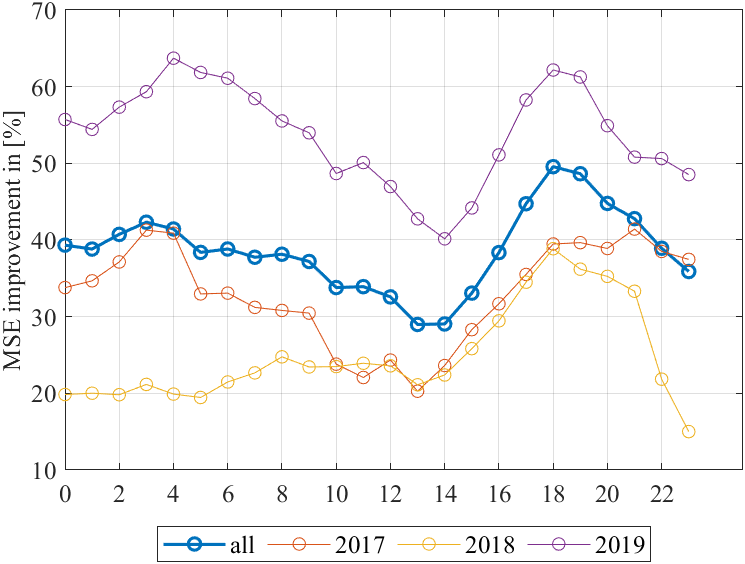}}
    \subfigure{\includegraphics[width=0.49\linewidth]{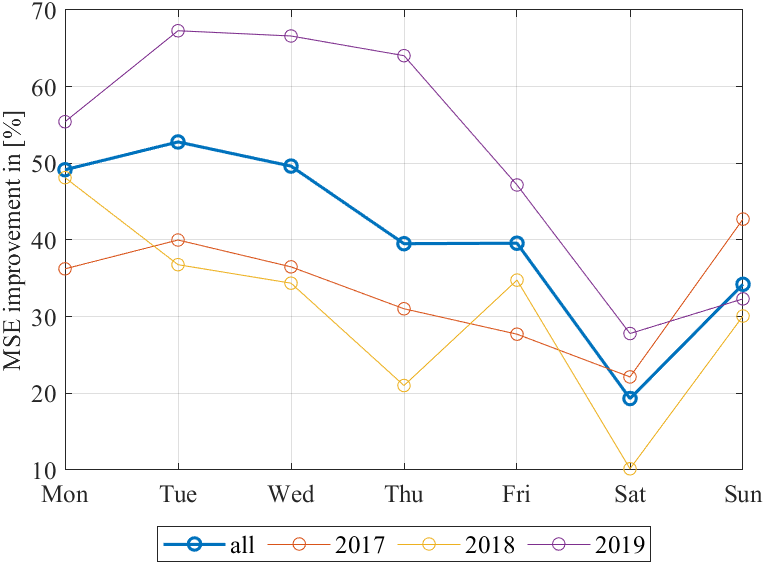}}
\caption{Average percentage MSE improvement for the day-ahead load forecast for each hour of a day (left) and for each weekday (right).}
\label{im_load_avmeanweekdaywise}
\end{figure}

To better attribute and understand the effect of load improvement on price, we also determine the percentage improvement in MSE for the hours of a day, and the days of the week, as shown in Figure \ref{im_load_avmeanweekdaywise}. The observed daytime and weekday structures in the TSO load forecast error are also evident in the improvement. During the day, hours 2 through 5 and 16 through 20 achieve the most considerable percentage improvement. Weekdays can be improved more than weekends; Tuesdays and Wednesdays show an especially strong improvement. In the TSO load forecast, these are the hours and days that have the largest mean error. Therefore, hours and days that have a sizeable mean error are the ones that have the most potential for improvement. Enhancing the load forecast by reducing this error is the primary goal of modelling and predicting the error of the TSO load forecast.  

\subsection{Impact of Improved Load Data on an Energy System Model}

In the previous Section\ref{subsec:ImprovedLoadData}, we proved that with a relatively straightforward approach, the ENTSO-E load data can be significantly improved. Thus, this approach is particularly suitable for energy system modellers to enhance critical input data. In the following, we quantify the impact of the improved load forecast on day-ahead wholesale price forecasts based on the \emph{em.power dispatch} model. To do this, we run the model twice, first using the original TSO-based load forecasts $\hat{L}$ and second, using the improved load forecasts $\hat{L}^*$ presented in Section\ref{subsec:ImprovedLoadData}. For both cases, we derive estimates of the day-ahead wholesale prices and calculate error measures comparing the results to actual observed day-ahead prices.

Using the improved load data set, we see an overall reduction in the error of the price estimator. For the entire time horizon, Table \ref{table_errormeasures_price_allyears} states a reduction of the MSE by 1.75~\%, the RMSE by 0.88~\% and the MAE by 0.42~\%. 

Comparing our results with those of other models in the same modelling class, we find that our model generates very good price estimates. \cite{Qussous2022}, for example, report an MAE of 9.44~€$/MWh$ for 2017, 8.88~€$/MWh$ for 2018 and 6.69~€$/MWh$ for 2019.

\begin{table}[H]
	\begin{center}
	\small
		\begin{tabular}{|l||r|r|r|r|r|r|r|r|}
			\hline
			\multicolumn{1}{|c||}{ } &
			\multicolumn{2}{c|}{all years} &
			\multicolumn{2}{c|}{2017} &
			\multicolumn{2}{c|}{2018} &
			\multicolumn{2}{c|}{2019} \\
			\hline
			 & Impr. & Orig & Impr & Orig & Impr & Orig & Impr & Orig  \\
			\hline
			[1] MSE  & 89.15 & 90.73 & 133.13 & 135.29 & 72.47 & 73.22 & 61.84 & 63.70 \\
			\hline
			[2] RMSE  & 9.44 & 9.53 & 11.54 & 11.63 & 8.51 & 8.56 & 7.86 & 7.98 \\
			\hline
			[3] MAE & 5.94 & 5.96 & 6.75 & 6.80 & 5.98 & 6.04 & 5.09 & 5.28 \\
			\hline
			Reduction [1] & 1.75\% & & 1.62\% & & 1.04\% &  & 3.01\% & \\
			\hline
			Reduction [2] & 0.88\% &  & 0.81\% &  & 0.52\% & & 1.49\% &  \\
			\hline
			Reduction [3] & 0.42\% &  & 0.85\% & & 1.02\% & & 3.65\% &  \\
			\hline
			
		\end{tabular}
		\bigskip
		\caption{Error measures for the price estimator of the \emph{em.power dispatch} model comparing the improved load forecasts (Impr.) by original load forecasts (Orig), given in [$MWh^2$] for MSE, in [$MWh$] for RMSE and MAE.}
		
		\label{table_errormeasures_price_allyears}
	\end{center}
\end{table}

Table \ref{table_errormeasures_price_allyears} further shows disaggregated error measures by year. It can be seen that an improvement in the error measure is achieved in all three years. However, the magnitude of this improvement varies; the relative error reduction is largest in 2019 and smallest in 2018. This observation correlates with the magnitude of the annual improvement in the load forecast, shown in Figure \ref{im_load_avmeanweekdaywise}.

\begin{figure}[H]
\centering
    \subfigure{\includegraphics[width=0.49\textwidth]{Load_improvement_MSE_in_perc_hourwise_eachyear.png}}
    \subfigure{\includegraphics[width=0.49\textwidth]{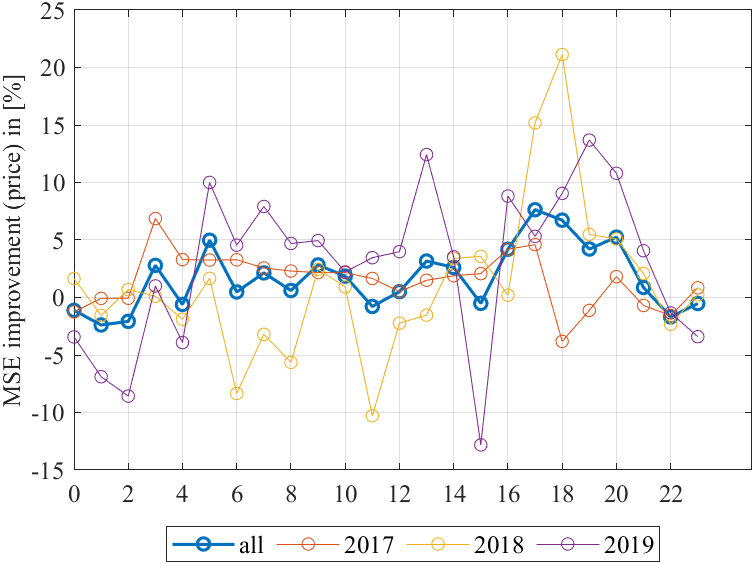}}
\caption{Average percentage MSE improvement of day-ahead load prediction (left) and day-ahead price estimators (right) for each hour of a day.}
\label{fig_hourly price and load}
\end{figure}

Furthermore, we analysed whether the improvement of the load estimator and the price estimator correlate with the hour of the day. Figure \ref{fig_hourly price and load} shows the average percentage improvement of the MSE of the day-ahead load prediction per hour of the day (left) and of the day-ahead price estimators (right). It can be seen that an hour's load and hour's price improvement do not correlate. Depending on the respective hour of the day, improvement of load prediction seems to have a different impact on the resulting price estimator. 

The reasoning for this discrepancy is two-fold: i) the model is more sensitive in one hour than in another hour, depending on the respective position in the merit order, and ii) an improvement in the load forecast in one hour may affect another hour due to temporal interdependencies such as storage operation and unit commitment decisions. 

\begin{figure}[H]
    \centering
    \includegraphics[width=1\linewidth]{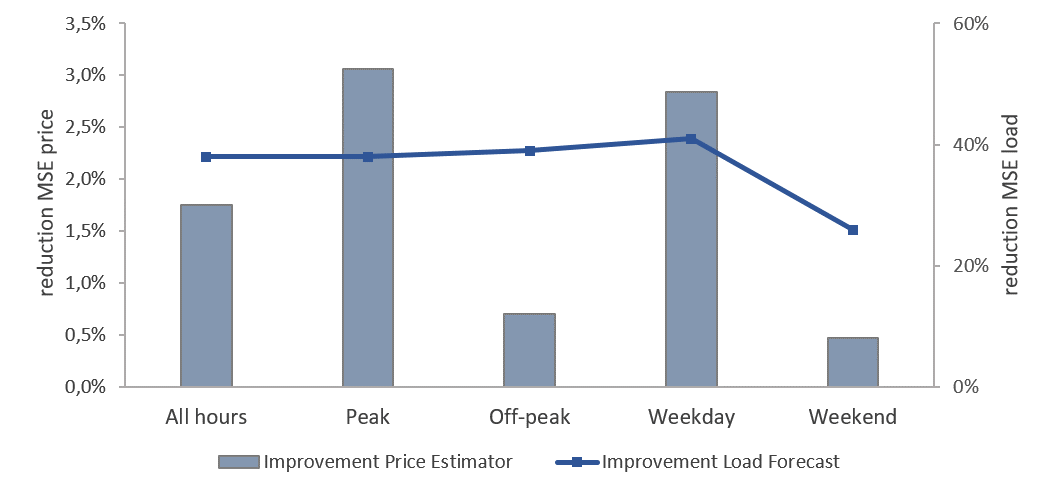}
    \caption{Percentage error reduction of the price estimator and the load in different time periods.}
    \label{figure_MSE_peak_offpeak}
\end{figure}

Having shown that the impact of better load forecasts on price forecasts derived in an energy system model is positive on average but varies between hours, we now examine the extent of error reduction at different points in time, starting with differentiation between high (Peak) and low demand (Off-Peak) periods. 
Figure \ref{figure_MSE_peak_offpeak} states the error reduction of the price estimator and that of the load forecast for the entire time period and time categories peak, off-peak, weekdays and weekend days. The most considerable error reduction of the price estimator is observed in peak hours and on weekdays in general. In the hours between 8 p.m. and 8 a.m. as well as on weekends, the effect on the price estimator is relatively low. On weekends, this observation correlates with the improvement of the load data, both of which are at their minimum. However, in off-peak hours, the impact on the price estimator is negligible, despite the great improvement in the load forecast.

As such, the model benefits significantly from improved load input data during peak hours and in total on weekdays, where demand and price levels are generally higher than off-peak hours and especially on weekends.

\FloatBarrier

\begin{figure}[H]
    \centering
    \includegraphics[width=0.9\linewidth]{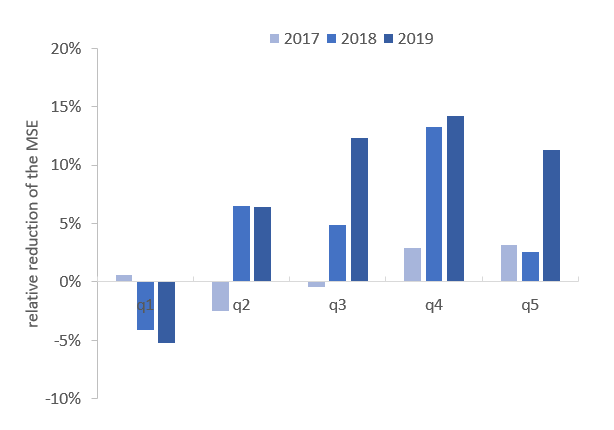}
    \caption{Relative error reduction of the price estimator in different price segments of the respective year from 2017 - 2019, starting with the lowest 20~\% quantile of electricity prices (q1) to the highest 20~\% quantile (q5).}
    \label{figure_MSE_price_quantiles}
\end{figure}

Based on the observation that price forecasts improve more during peak periods than in off-peak periods, we analyse the relation between wholesale price and forecast improvement. Figure \ref{figure_MSE_price_quantiles} shows the improvement of the price estimator for five different price segments where electricity prices are equally separated in 20~\% quantiles based on their level. The first quantile (q1) represents the lowest 20~\% quantile and the last quantile (q5) the highest 20~\% quantile of electricity prices of the respective year between 2017 and 2019.

It can be seen that the error reduction of the price estimator is most relevant in hours with high and medium prices. Overall, the largest improvement can be observed in 2018 and 2019 with an MSE reduction of nearly 15~\%, here at times with the 60-80~\% highest prices. In contrast, the improved load forecast data does not lead to a better price estimator in low-price periods. In all years, we even observe an increasing error in these price ranges. In summary, the improved load forecast is most beneficial for the model in the hours when the market equilibrium is found on the right side in the merit order, i.e., where changes or errors in the demand have the highest price impact.

\FloatBarrier

Hence, our analysis shows that the price forecasts are generally better when a) demand is high and b) prices are high. As traded volumes (in monetary terms) are the product of prices and volumes, it is interesting to note that price forecast improvement is highest when it matters the most.

\section{Conclusion}
\label{sec:conclusion}

This paper discusses data preprocessing in the context of fundamental energy system models. We present a simple time series model to improve the TSO-based load forecast data provided by ENTSO-E. The model captures and removes systematic biases and autoregressive structures present in the load forecast errors. Since the model is applied to observed forecast errors rather than to the load data itself and does not include load-specific external variables, it can be easily transferred to the preprocessing of other quantities of interest. 

To analyse the effect of enhanced load forecasts on electricity system models, we feed the improved load forecast data into the \emph{em.power dispatch} model. The model is used to generate price estimates for the German day-ahead electricity market, and we present the structure, assumptions, and optimisation equations of the model in detail. Concerning the effect of sequentially preprocessed inputs, we find that the benefits of sequentially improved load forecasts strongly depend on the respective price level, with more extensive benefits for higher price levels. This is a universal result in line with fundamental theory since in merit order markets, the impact of load changes on price changes increases with the overall level. We find that in phases of relatively high prices, as in 2018 and 2019, the continuous and sequential, i.e., day-by-day, load data preprocessing leads to an average reduction of \emph{em.power dispatch} 's prices forecast mean squared error by nearly 15~\%. Hence, as the value of traded energy is the product of prices and volumes, our analysis shows that forecasts are generally better when a) demand is high and b) prices are high, i.e., when it matters the most.

Based on these findings, we recommend energy system modellers to carefully analyse not only the structure and equations of their models but also the quality of input data. This paper demonstrated in the empirical setting of the German wholesale electrictiy market that input data can be improved significantly and that these improvements can be achieved with very simple time-series models. Furthermore, we demonstrate that results of the energy system model benefit from the improved input data. 

\section{Data availability}
Datasets related to this article and a source code for the entire project are available in a public GitHub repository. On \hyperlink{https://github.com/ProKoMoProject/Enhancing-Energy-System-Models-Using-Better-Load-Forecasts}{github.com/ProKoMoProject/Enhancing-Energy-System-Models-Using-Better-Load-Forecasts}. you find code and data for the time series model improving the load forecasts as well as code and data for the energy system model. The codes reproduce the benchmarks from the paper.

\section*{Acknowledgements}
The work was supported by the German Federal Ministry of Economic Affairs and Climate Action through the research project "ProKoMo - Better price forecasts in the energy sector by combining fundamental and stochastic models" within the Systems Analysis Research Network of the 6th energy research program. 

\bibliography{ProKoMo}
\newpage

\appendix
\section{Descriptive statistics for various time resolutions}

\begin{table}[h]
	\begin{center}
	\footnotesize
	%\scriptsize
		\begin{tabular}{|l||r|r|r|r|r|r|r|}
			\hline
				Day &	Mean	&	Median	&	Minimum	&	Maximum	&	5\%-Q.	&	95\%-Q.	&	Std.	\\
			\hline
                Mon.	&	1,517.71	&	1,436.63	&	-6,277.00	&	12,930.75	&	-1,895.75	&	5,111.45	&	2,159.26	\\
                Tue.	&	1,685.17	&	1,581.50	&	-4,867.25	&	11,469.00	&	-1,473.05	&	5,126.95	&	1,984.32	\\
                Wed.	&	1,601.81	&	1,516.00	&	-7,868.50	&	9,053.75	&	-1,242.65	&	4,667.90	&	1,874.89	\\
                Thu.	&	1,361.43	&	1,365.25	&	-17,543.50	&	9,520.25	&	-1,372.88	&	4,330.47	&	1,896.08	\\
                Fri.	&	944.31	&	990.75	&	-10,596.88	&	9,772.25	&	-2,060.85	&	3,911.48	&	1,892.38	\\
                Sat.	&	-102.26	&	-43.63	&	-7,661.75	&	7,809.00	&	-2,967.80	&	2,380.13	&	1,696.76	\\
                Sun.	&	-378.19	&	-340.13	&	-7,752.50	&	7,522.00	&	-3,378.53	&	2,458.15	&	1,859.72	\\
            
			\hline
			
		\end{tabular}
		\caption{Weekday wise averaged descriptive statistics of TSO load forecast errors for the years 2016 to 2019. All variables are given in [$MWh$].}
		
		\label{table_descriptives_load_allyears_weekday}
	\end{center}
\end{table}

\begin{table}[H]
	\begin{center}
	\footnotesize
		\begin{tabular}{|r||r|r|r|r|r|r|r|}
			\hline
			Hour	&	Mean	& Median & 	Minimum	&	Maximum	&	5\%-Q.	&	95\%-Q.	&	Std.	\\
			\hline
            1	&	699.64	&	742.75	&	-5,710.00	&	6,594.75	&	-2,270.28	&	3,509.15	&	1,713.19	\\
            2	&	652.07	&	696.75	&	-6,302.25	&	6,701.50	&	-2,290.58	&	3,460.69	&	1,711.79	\\
            3	&	646.41	&	657.75	&	-6,558.25	&	6,882.75	&	-2,246.04	&	3,535.18	&	1,739.51	\\
            4	&	6,90.28	&	687.50	&	-8,539.00	&	6,888.25	&	-2,269.38	&	3,635.30	&	1,760.22	\\
            5	&	855.45	&	852.75	&	-15,353.75	&	7,617.25	&	-2,298.63	&	3,959.94	&	1,991.04	\\
            6	&	943.53	&	1,032.50	&	-17,543.50	&	9,113.00	&	-2,658.48	&	4,335.81	&	2,295.79	\\
            7	&	946.49	&	1,075.75	&	-20,358.00	&	11,469.00	&	-2,896.14	&	4,688.48	&	2,515.40	\\
            8	&	912.14	&	1,011.00	&	-19,681.75	&	11,152.25	&	-2,795.23	&	4,529.93	&	2,461.27	\\
            9	&	908.80	&	959.00	&	-16,540.25	&	11,658.25	&	-2,692.56	&	4,594.01	&	2,387.68	\\
            10	&	902.67	&	900.00	&	-15,234.63	&	12,930.75	&	-2,836.34	&	4,707.63	&	2,417.03	\\
            11	&	916.34	&	926.75	&	-14,149.88	&	10,484.25	&	-2,791.68	&	4,646.26	&	2,401.02	\\
            12	&	923.30	&	951.00	&	-14,261.88	&	10,471.25	&	-2,818.44	&	4,782.05	&	2,414.55	\\
            13	&	907.81	&	982.00	&	-15,839.88	&	9,777.25	&	-2,980.00	&	4,740.98	&	2,446.60	\\
            14	&	785.04	&	882.75	&	-15,116.63	&	9,306.75	&	-2,963.74	&	4,549.70	&	2,412.79	\\
            15	&	718.73	&	807.50	&	-15,220.38	&	9,510.50	&	-2,933.24	&	4,477.88	&	2,396.13	\\
            16	&	900.33	&	895.50	&	-14,434.38	&	9,520.25	&	-2,777.48	&	4,756.91	&	2,323.86	\\
            17	&	1,080.73	&	1,051.50	&	-13,068.00	&	9,951.25	&	-2,526.88	&	4,832.60	&	2,285.24	\\
            18	&	1,172.53	&	1,128.00	&	-12,133.38	&	10,094.50	&	-2,110.33	&	4,819.08	&	2,140.34	\\
            19	&	1,180.48	&	1,161.25	&	-11,652.75	&	9,409.75	&	-1,985.76	&	4,610.81	&	2,078.72	\\
            20	&	1,061.99	&	1,071.00	&	-9,940.75	&	8,430.50	&	-1,956.98	&	4,248.70	&	1,978.27	\\
            21	&	875.01	&	909.25	&	-8,443.13	&	7,422.00	&	-2,121.79	&	3,943.60	&	1,888.30	\\
            22	&	877.85	&	872.75	&	-6,668.50	&	6,932.25	&	-2,032.00	&	3,853.51	&	1,821.24	\\
            23	&	863.47	&	806.00	&	-5,642.38	&	6,947.25	&	-1,928.34	&	3,706.63	&	1,726.26	\\
            24	&	729.88	&	763.25	&	-5,607.88	&	7,011.50	&	-2,262.01	&	3,560.98	&	1,701.36	\\

			\hline
			
		\end{tabular}
		\caption{Hourly averaged descriptive statistics of TSO load forecast errors for the years 2016 to 2019. All variables are given in [$MWh$].}
		
		\label{table_descriptives_load_allyears_hourly}
	\end{center}
\end{table}

\newpage
\section{Nomenclature}
\label{Nomenclature}
\makenomenclature
\mbox{}
\nomenclature[A]{$\mathit{BP}$}{\hspace{2.9em} Time blocks for primary control power}
\nomenclature[A]{$\mathit{BS}$}{\hspace{2.9em} Time blocks for secondary  control power}
\nomenclature[A]{$I$}{\hspace{3em} Electricity generation capacity cluster}
\nomenclature[A]{$N,\mathit{NN}$}{\hspace{3em} Node}
\nomenclature[A]{$\mathit{STM}(I)$}{\hspace{2.7em} Mid-term storage [Subset of I]}
\nomenclature[A]{$\mathit{STL}(I)$}{\hspace{3em} Long-term storage [Subset of I]}
\nomenclature[A]{$\mathit{RES}(I)$}{\hspace{3em} Intermittent renewables [Subset of I]}
\nomenclature[A]{$T$}{\hspace{3em} Time steps}
\nomenclature[A]{$\mathit{tfirst}(T)$}{\hspace{2.5em} First hour of a rolling window}
\nomenclature[A]{$\mathit{tlast}(T)$}{\hspace{2.6em} Last hour of a rolling window}

\nomenclature[B]{$\eta_i$}{\hspace{3em} Efficiency rate of a generation technology}
\nomenclature[B]{$\mathit{af}_{i,n,t}$}{\hspace{3em} Availability factor for generation capacities}
\nomenclature[B]{$\mathit{cap}_{i,n,t}$}{\hspace{3em} Installed generation capacity [$MW_{el}$]}
\nomenclature[B]{$\mathit{chp}_{i,n,t}$}{\hspace{3em} Minimum electricity output of combined heat power units [$MWh_{el}/h$]}
\nomenclature[B]{$\mathit{curtc}$}{\hspace{3em} Costs for RES curtailment [€$/MWh_{el}$]}
\nomenclature[B]{$\mathit{epf}$}{\hspace{3em} Energy-power factor for mid-term storage plants [$MWh_{el}/MW_{el}$]}
\nomenclature[B]{$d_{n,t}$}{\hspace{3em} Electricity demand [$MWh_{el}/h$]}
\nomenclature[B]{$\mathit{g}_{i,n,t}^{min}$}{\hspace{3em} Minimum generation of a running unit}
\nomenclature[B]{$\mathit{ntc}_{n,nn,t}$}{\hspace{2.9em} Net transfer capacities [$MWh_{el}/h$]}
\nomenclature[B]{$\mathit{out}_{i,n,t}$}{\hspace{3em} Power plant outages [$MW_{el}$]}
\nomenclature[B]{$\mathit{sc}_{i,n,t}$}{\hspace{3em} start-up costs [€$/MW_{el}$]}
\nomenclature[B]{$\mathit{vc}_{i,n,t}^{FL}$}{\hspace{3em} Variable generation costs at full load [€$/MWh_{el}$]}
\nomenclature[B]{$\mathit{vc}_{i,n,t}^{ML}$}{\hspace{3em} Variable generation costs at minimum load [€$/MWh_{el}$]}
\nomenclature[B]{$\mathit{voll}$}{\hspace{3em} Value of lost load [€$/MWh_{el}$]}
\nomenclature[B]{$\mathit{wv}_{i,n,t}$}{\hspace{3em} Water value for hydro reservoirs and long-term storage [€$/MWh_{el}$]}
\nomenclature[C]{$\mathit{CURT}_{i,n,t}$}{\hspace{2em} RES curtailment [$MWh_{el}/h$]}
\nomenclature[C]{$\mathit{CL}_{i,n,t}$}{\hspace{3em} Charging activity for long-term storage [$MWh_{el}/h$]}
\nomenclature[C]{$\mathit{CM}_{i,n,t}$}{\hspace{3em} Charging activity for mid-term storage [$MWh_{el}/h$]}
\nomenclature[C]{$\mathit{FLOW}_{n,nn,t}$}{\hspace{1.5em} Electricity flow from node n to node nn [$MWh_{el}/h$]}
\nomenclature[C]{$G_{i,n,t}$}{\hspace{3em} Electricity generation [$MWh_{el}/h$]}
\nomenclature[C]{$P_{i,n,t}^{on}$}{\hspace{3em} Running capacity [$MW_{el}$]}
\nomenclature[C]{$\mathit{PCR}_{i,n,t}$}{\hspace{3em} Primary control reserve [$MW_{el}$]}
\nomenclature[C]{$\mathit{SCR}_{i,n,t}^{pos}$}{\hspace{2.8em} Positive secondary control reserve [$MW_{el}$]}
\nomenclature[C]{$\mathit{SCR}_{i,n,t}^{neg}$}{\hspace{2.8em} Negative secondary control reserve [$MW_{el}$]}
\nomenclature[C]{$\mathit{SHED}_{n,t}$}{\hspace{2.5em} Load shedding [$MWh_{el}/h$]}
\nomenclature[C]{$\mathit{SL}_{i,n,t}$}{\hspace{3em} Storage level of PSP [$MWh_{el}$]}
\nomenclature[C]{$\mathit{SU}_{i,n,t}$}{\hspace{3em} Start-up activity of a generation unit [$MW_{el}$]}
\nomenclature[C]{$\mathit{TC}$}{\hspace{3em} Total system costs [€]}

%\printnomenclature

\begin{figure}[H]
    \centering
    \includegraphics[width=1.2\linewidth]{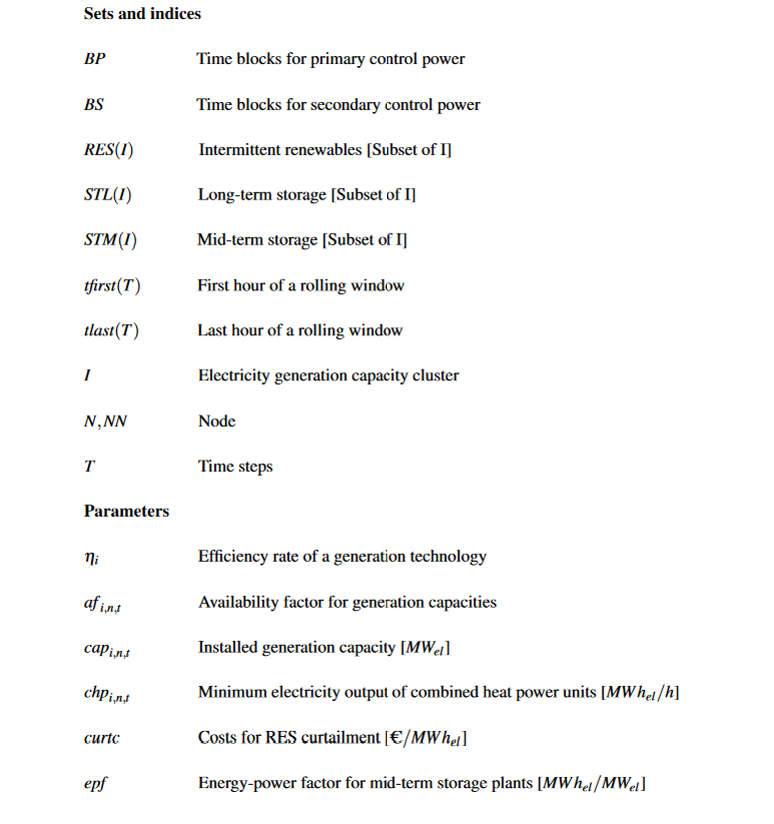}
%    \caption{Percentage error reduction of the price estimator and the load in different time periods.}
%    \label{figure_MSE_peak_offpeak}
\end{figure}
\FloatBarrier
\newpage

\begin{figure}[H]
    \centering
    \includegraphics[width=1.2\linewidth]{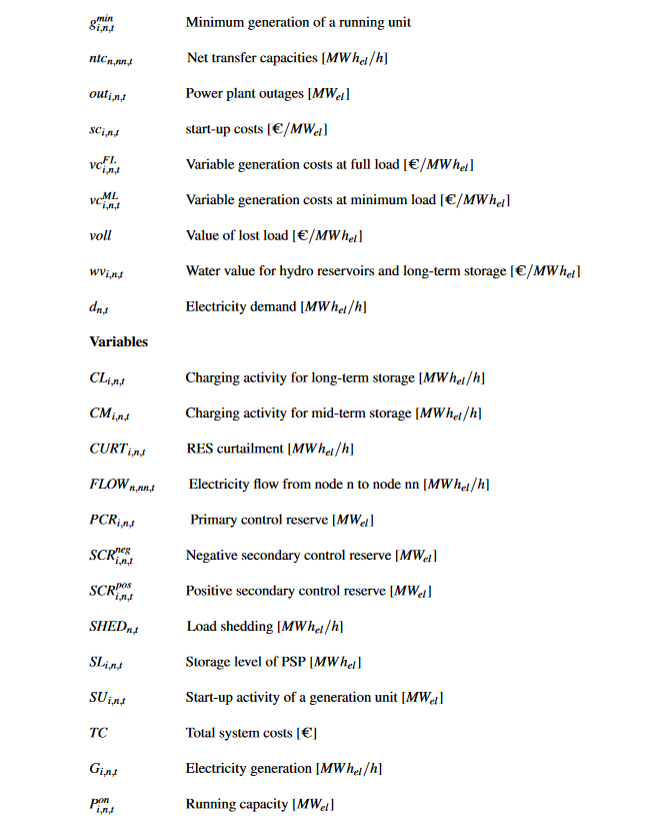}
%    \caption{Percentage error reduction of the price estimator and the load in different time periods.}
%    \label{figure_MSE_peak_offpeak}
\end{figure}
\FloatBarrier

\newpage
\section{Error measures for different rolling window lengths}

\begin{table}[ht]
	\begin{center}
	\footnotesize
		\begin{tabular}{|r||r|r|r|r|r|r|r|r|}
			\hline
				&	all	&	2017	&	2018	&	2019	\\
			\hline
            MSE	&	4,948,990.80	&	2,272,350.07	&	4,575,637.58	&	3,133,176.51	\\
            RMSE	&	2,224.63	&	1,507.43	&	2,139.07	&	1,770.08	\\
            MAE	&	1,691.37	&	1,132.86	&	1,462.48	&	1,377.82	\\
			\hline
		\end{tabular}
		\caption{Error measures (MSE, RMSE, MAE) for the the improved day ahead load forecast with a rolling window length of three month. MSE is given in [$MWh^2$], RMSE and MAE in [$MWh$].}
		\label{table_errormeasures_3mRW_allyears}
	\end{center}
\end{table}

\begin{table}[h]
	\begin{center}
	\footnotesize
		\begin{tabular}{|l||r|r|r|r|r|r|r|r|}
			\hline
				&	all	&	2017	&	2018	&	2019	\\
			\hline
            MSE	&	4,016,010.14	&	4,896,546.88	&	4,355,814.17	&	2,795,669.38	\\
            RMSE	&	2,004.00	&	2,212.81	&	2,087.06	&	1,672.03	\\
            MAE	&	1,304.12	&	1,200.16	&	1,411.57	&	1,300.63	\\
			\hline
			
		\end{tabular}
		\caption{Error measures (MSE, RMSE, MAE) for the the improved day ahead load forecast with a rolling window length of six month. MSE is given in [$MWh^2$], RMSE and MAE in [$MWh$].}
		
		\label{table_errormeasures_6mRW_allyears}
	\end{center}
\end{table}

%\newpage

\end{document}